# APPLYING SOFTWARE ENGINEERING SOLUTIONS TO LAW FIRM MANAGEMENT, NIGERIA AS A CASE STUDY

Chinonyerem Eleweke  and Kazeem Oluwakemi Oseni

School of Computer Science and Technology, University of Bedfordshire, UK

## ABSTRACT

*Legal technology has changed the way law firms are managed worldwide. Substantial research has been undertaken on the role of legal technology in law firm management especially in developed countries. Though, most studies have only focused on the benefits and challenges, and have failed to analyse law firm management areas requiring software solutions.*

*The principal objective of this paper was to investigate the level of technology adoption among Nigerian law firms, as well as to develop a software solution to automate work processes in identified areas. This investigation was done using systematic literature review to gather relevant data on the subject area and identify knowledge gaps.*

*Findings from the research indicated a need for further analysis of the various areas in law practice that could require software solutions. The findings also discussed the implementation of a property management module which is an important contribution to the management of law firms in Nigeria. A speech-to-text transcription feature was also implemented to eliminate the need for lengthy typing.*

## 1. INTRODUCTION

There is no gainsaying that software engineering is a key driver of growth in the digital technology industry as many technological innovations are dependent on it [1]. It is therefore fundamental to technological advancement in the legal sector. Moreover, legal technology has received substantial scholarly consideration in recent years with the majority focusing on developed countries. There are several definitions of legal technology, however, three definitions will be adopted in this study.Statista [2] defined Legal Technology (Legal Tech) as the application of software solutions and technological innovations to assist in the provision of services in the legal sector. According to Whalen [3], legal tech can also be referred as any device that can be used to enable users to interact with the law. It comprises of technologies that can be used for legal purposes.

Additionally, legal tech is the automation of all aspects of legal practice with the use of Information and Communication Technology (ICT), in order to aid legal service providers; and provide legal advice to clients[4].It has been established that technology has changed the way people in the legal profession carry out daily activities by making them easier and faster, thereby making law firm staff more productive[5]. Legal technology has also improved the efficiency of law firms worldwide[6]. However, despite these benefits, Nigerian law firms have been slow to adopt technology. Recent studies[7], [8], [9], [10] shows that challenges such as poor power supply, poor internet connection, and lack of technical knowledge, and so on, have greatly hindered the adoption of technology in Nigerian law firms.

On the other hand, in spite of these recent findings about the role of technology in legal practice, far too little attention has been paid to the involvement of legal practitioners in real estate property management. There is a need for an in-depth analysis of this area as it would greatly improve existing legal tech solutions.This paper aims to analyse the current situation in Nigerian law firms concerning technology adoption. In doing this, it would identify the research gaps in Nigerian legal technology; and contribute to the body of knowledge by developing solutions that address the identified gaps.

## 2. LITERATURE REVIEW

### 2.1. OVERVIEW

In recent years, there has been an increasing amount of literature on Legal Technology. However, they are mostly focused on developed countries. As discussed by Statista[2], the COVID-19 pandemic played a major role in technology adoption in the legal sector worldwide. It was noted that, as a result of controlled movement, technology was adopted to foster remote collaboration in various legal sectors, including courts.

Similarly, since people could no longer go out as usual, law firms needed to find ways to continue serving their clients. Courts also needed to continue conducting hearings virtually; hence the adoption of video conferencing technologies. For example, Zoom is one of the tools that can be used to hold meetings with multiple attendees. As noted by Aidonojie et al.[10], during the COVID-19 pandemic, the use of technology was widely adopted by law firms in Nigeria to reduce human contact while also making sure to continue the smooth running of daily activities. Another major driver for technology adoption among legal practitioners has been the need to develop new, cost-effective ways of carrying out tasks, which helps them stand out among competitors [2]. Law firms need to continually finding ways to stay relevant, or they risk losing high-paying clients to better-performing firms. Clients would likely prefer to use firms that offer valuable services, like automated case progress tracking for instance.

Additionally, clients' requirements for timely reports on the progress of their cases, the ability to view their case files, and to connect with their lawyers virtually led many law firms to embrace technology[11]. Clients may want to receive updates on their cases without physically visiting the law firm. For example, clients who live several miles away from the firm would prefer to access information on their case progress remotely rather than travel for hours to visit the firm. In a recent study by Statista[12], a large number of legal tech startup companies have emerged all over the world and this number is expected to keep growing till 2027 and beyond. However, very few of these companies exist in Nigeria and Africa as a whole. This is a clear indication that there is still work to be done in the Nigerian legal tech industry.

Given all that has been mentioned so far, it is evident that the COVID-19 pandemic was a significant catalyst in the adoption of technology by law firms. This suggests that work processes in most law firms were carried out manually before the pandemic. It is also noteworthy that most of the advancements in legal technology have been solely adopted in developed countries. Hence, there is a need for growth and innovation in the African legal tech industry, particularly in Nigeria.

## 2.2. Benefits of Legal Technology

As previously discussed, technology adoption has recently become widespread in the legal sector. There are several benefits associated with this adoption. As noted by Owoeye and Mabawonku[13], access to Information and Communication Technology (ICT), has led to a major improvement in the practice of law globally.

One benefit of technology is that it has helped to ease the large volume of document processing and storage associated with law practice[14]. Document management is important in any organisation as it makes the storage and retrieval of documents easier and faster. For instance, using Microsoft Office to type and store files on computers makes it easier to access such documents. In comparison to hardcopy documents which can easily get damaged, storing documents on computers or the cloud makes it accessible and reduces the risk of loss or damage. Additionally, technology helped legal practitioners and law firms to complete tasks easily and quickly, thereby becoming more productive[5]. Tasks like time management, for example, can be very tedious when done manually. However, with automated time management software, effort can be put into more important tasks, which would greatly increase productivity.

Thirdly, clients are able to monitor the progress of their lawsuits without physically visiting their lawyers [5]. Clients being able to monitor their case progress remotely would encourage them to continue to patronise the law firm as it would save them the time and cost required to always visit whenever they require an update on their lawsuit. Also, technology has assisted in legal research which is a crucial part of legal education and practice [5]. Instead of having to borrow books from a library or purchase weekly law reports, lawyers can find whatever articles they need on the internet, which makes their research process easier and less time-consuming.

Furthermore, technology has changed the way records of court proceedings are taken from merely writing them down to recording them digitally and listening to them later on[15]. Court proceedings may be lengthy and writing them down would be time-consuming and inefficient. While recording may offer a better alternative, it still requires more time to listen to the recording after the court session.These points are consistent with data obtained from a study by Bloomberg[6] which showed that law firms worldwide benefit from legal tech in various ways including an increase in organisational efficiency and improved workflow. Figure 1 below is a bar chart representing the survey results.

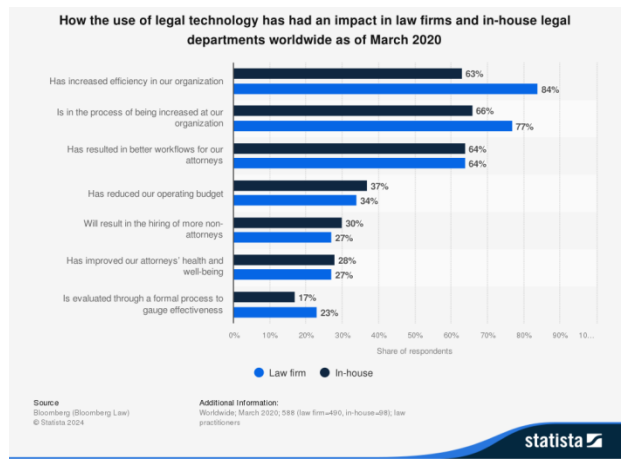

Figure 1: Survey on benefits of legal tech [6].

From Figure 1 above, it is apparent that a large number of lawyers (84%) attest to the fact that legal tech has led to an increase in their organisational efficiency. It can also be concluded that the second highest percentage (77%) noted that increased efficiency is in progress.

Similarly, 64% of respondents agreed that legal technology has led to an enhancement in lawyers' work process.However, very few lawyers agreed that legal tech increased their health, reduced their budgets, and required them to hire more legal staff.

Therefore, it can be deduced that technology has not only improved productivity and efficiency in law firms, but it has also led to improved client satisfaction and better legal research. However, some challenges may hinder technology adoption among Nigerian lawyers. These challenges will be discussed in detail in the next section.

## 2.3. Challenges Hindering Legal Tech Adoption in Nigeria

Despite the benefits identified in the previous section, there are many hindrances to technology adoption, especially in Nigeria. According to Damkor, Joro & Manasseh[15], various issues hinder legal practitioners from fully embracing technology.Existing literature on technology and Nigerian law firms pays particular attention to the challenges faced by law firms which prevent them from adopting technology. One major challenge is unreliable internet access[9], [10]. Internet connection is a basic requirement for the use of most modern technology solutions; therefore, it is imperative that law firms need to be connected to the internet before they can adopt the majority of the available legal technology. A notable example is electronic mail (email), and without an internet connection, it is impossible to send an email, let alone other sophisticated technology.

Secondly, there is the issue of an unstable supply of electricity [9], [10]. Most devices need power supply to function. A printer, for instance, cannot run if there is no power source. This could lead to lawyers being unable to print documents stored on their computer, and set-backs in the workflow. Therefore, without stable electricity, law firms would not be able to adequately utilise technology.Another issue as noted by Paul & Uko[7], is that Nigerian lawyers are not technologically proficient and depend solely on hardcopy documents. It is difficult to adopt technology if lawyers who are the main law firm staff do not know how to use them. This issue could be addressed by ensuring mandatory training for all law firm staff[16]. However, a broader perspective was adopted by San, Mohamad and Sule[16] who argued that the issues faced by legal practitioners include, but are not limited to case management, law firm management, research, document processing, client interaction and working with fellow lawyers. Lawyers need a way to efficiently manage their law firms including their clients, cases, and documents. They also need an easier way to conduct legal research and interact with colleagues.

Overall, these studies consistently indicate that while poor internet connection and power supply are major issues hindering technology adoption, law firms also struggle with managing their documents, managing their law firm, managing clients and conducting research.This paper therefore builds on these issues and proposes a software solution that addresses them. The next section outlines proposed recommendations to address these issues.

## 2.4. Recommendations

With all the challenges previously identified, recent studies have proposed several solutions to tackle them.Firstly, Paul and Uko[7] suggest the need for relevant authorities to monitor the standard of internet services offered by internet service providers. This would be advantageous to both lawyers and the general public. If the Nigerian government sets a standard for providing internet services by network providers, these providers would be forced to improve their products or risk losing their business. This would in turn eliminate the issue of unstable internet.Secondly, the Nigerian Bar Association (NBA) ought to ensure that lawyers take part in seminars that educate them on the importance and use of technology in legal practice yearly. They could also consider making it a criterion for renewing practice permits by lawyers[16]. Educating lawyers would help improve their proficiency and encourage them to use technological tools that would enhance their productivity. Making these trainings mandatory on the other hand, would ensure that lawyers are up to date with technological trends and can collaborate among themselves.

Furthermore, Oweye and Mabawonku[13] suggest the development of a generic software accessible to all Nigerian law firms. The Nigerian Bar Association needs to work with the Nigerian Council of Legal Education to create a policy that can be adopted in the development of said software. Compared to custom software that may be costly to acquire, developing generic legal software would be less expensive which would mean that more firms would readily embrace it. Furthermore, the involvement of the relevant authorities, would ensure the software is made up to standard and meets users' needs.In addition to relevant authorities creating policies for legal tech development, legal practitioners also need to work together with software engineers to develop systems that can solve their business needs[7]. This collaboration would give developers a broader perspective on how best to create solutions that would be useful to both lawyers and non-lawyers in the legal sector.In view of all that has been mentioned so far, one may suppose that to find lasting solutions to the issues identified in the previous section, there is a need for collaboration among Nigerian lawyers, software engineers, as well as relevant authorities in the Nigerian legal sector. The quality of internet service provided in Nigeria needs to be regulated and lawyers need to be trained on the use of technology.

# 3. RESEARCH METHODOLOGY

Systematic literature review was adopted in this article to gain meaningful insights from existing literature on legal technology. Literature was identified by searching electronic databases covering the period 2019-2024. Databases searched include the University of Bedfordshire's Library catalogue, Google Scholar, IEEE Xplore, ACM Digital Library, International Journal of Law and Politics Studies, African Journal of Humanities, and ResearchGate.

## 3.1. USE CASE

A use case gives a description of a system's intended functionalities. It shows how various users interact with the system. The actors in the system are the law firm staff and clients. These actors can interact with the system in various ways, which would be discussed below.

## 3.2. Use Case Actors

The key actors in the proposed system are:

1. Clients: this represents clients that require legal services from a law firm.
2. Staff: represent the employees of a law firm and can either be lawyers or clerks.

## 3.3. Use Cases

1. Register and Login: every user in the system is able to register and login to the system.

2. Manage clients: staff are able to view a list of clients.
3. Manage cases: staff can create, view and update cases.
4. Manage court records: staff are able to input, view and update court records for each case.
5. View dashboard: users can view dashboard.
6. Manage appointments: users can create, update and cancel appointments.
7. Manage tenants: staff can add, update and view tenants.
8. Manage properties: staff can create, read, and update properties.
9. Manage transactions: staff can create new transactions, update transactions and view transactions.
10. View case progress: clients can view updates on their cases.
11. View report: both staff and clients can view reports.
12. View notifications: both staff and clients can view their notifications.

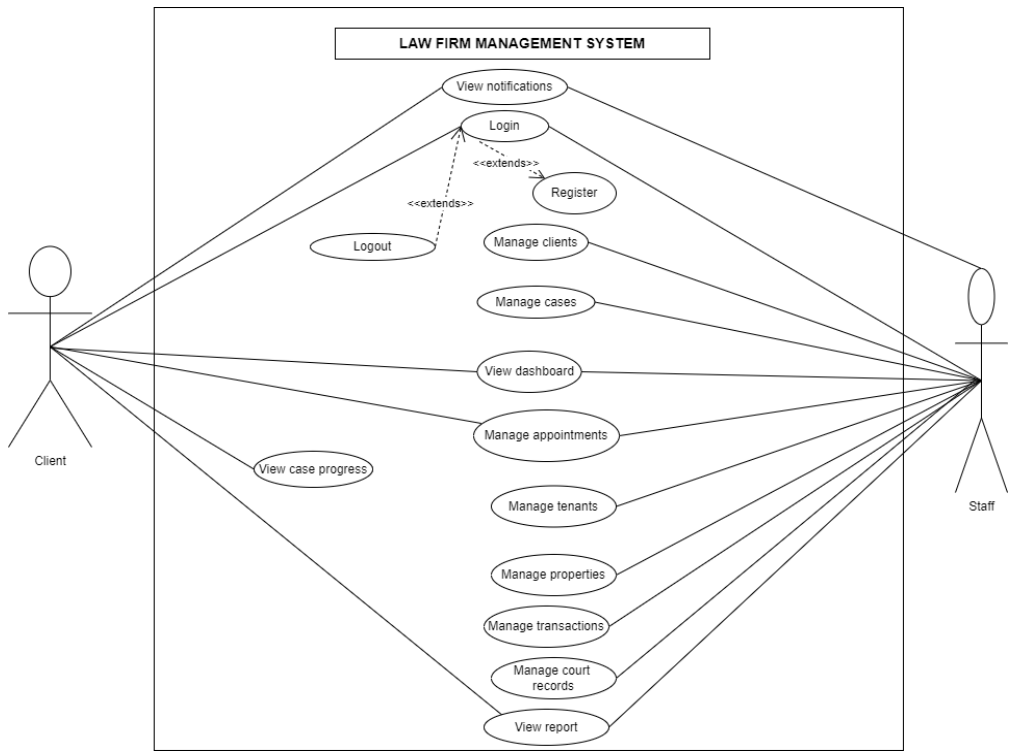

Figure 2: Use case diagram.

# 4. ANALYSIS OF FINDINGS

One of the aims of this study was to analyse the current use of technology in Nigerian law firms and to propose potential solutions to improve their work processes. In the literature review, existing research on legal technology was analysed.

Findings from the review indicated that many recent studies noted the importance of adopting legal tech in law firms; and the challenges of its adoption. However, very little was found in the existing literature about users' feedback on legal tech solutions in Nigeria and the legal practice areas that may require technology automation.

Based on the findings, software was developed to offer new features in addition to what is already available. A speech-to-text transcription feature, which solves the problem of typing for long hours was implemented. A property management module was also implemented to help lawyers manage properties seamlessly and generate reports for each property on demand. Figures3 and 4 below show the user interface of the speech-to-text and property management features respectively.

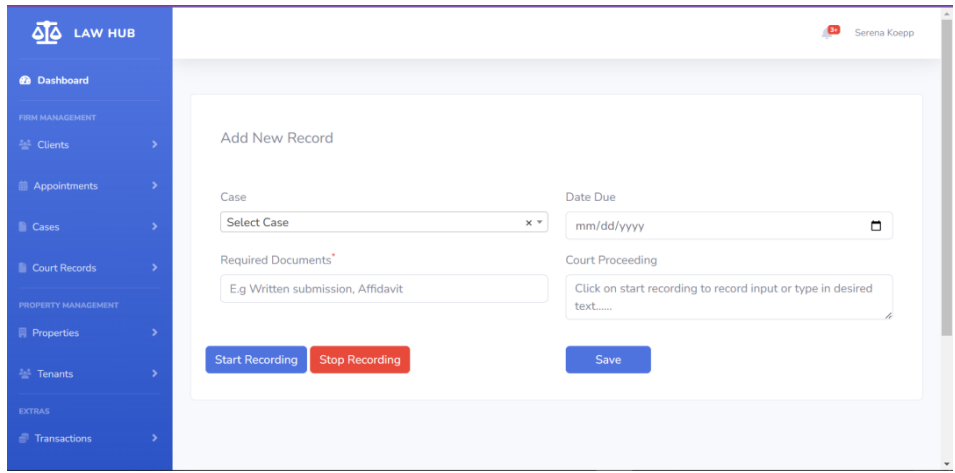

Figure 3: Speech-to-text feature

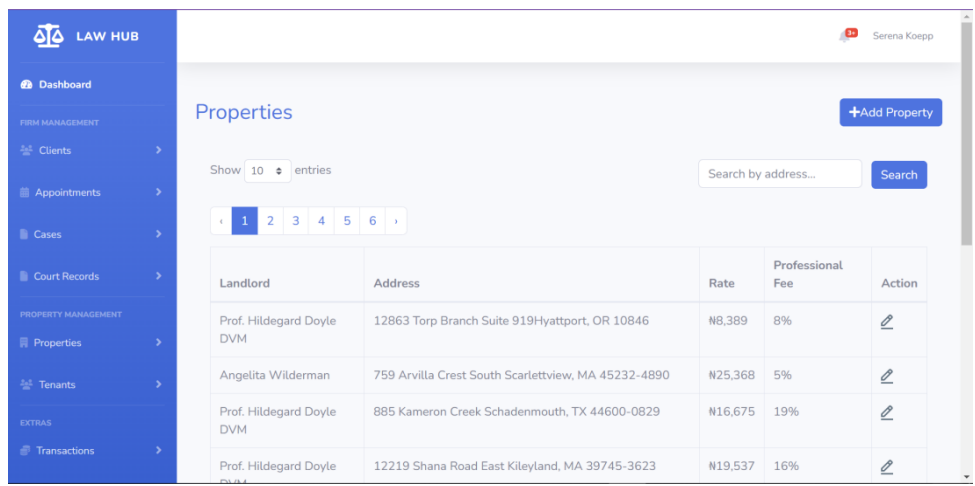

Figure 4: Property management feature

# 5. CONCLUSION

This paper has also been able to analyse the rate of technology adoption in Nigerian law firms and to develop an artefact to improve their work process. It was found that Nigerian law firms have been slow in adopting technology due to poor internet, electricity, and lack of technical proficiency. In conclusion, the findings and recommendations from this paper have provided a basis for continuous innovation in the Nigerian legal tech industry.